\documentclass[12pt,preprint]{aastex}

\shorttitle{The variable PWN of PSR B0540-69}
\shortauthors{A.De Luca et al.}

\begin{document}


\title{HST multi-epoch imaging of the PSR B0540-69 system unveils
a highly dynamic synchrotron 
nebula.\footnote{Based on observations with the NASA/ESA 
Hubble Space Telescope, obtained at the Space Telescope Science 
Institute, which is operated by AURA, Inc. under contract No. 
NAS 5-26555.}}


\author{A. De Luca}
\affil{INAF - Istituto di Astrofisica Spaziale e Fisica Cosmica,
Via Bassini 15, I-20133 Milano, Italy}
\email{deluca@iasf-milano.inaf.it}
\and 
\author{R.P. Mignani}
\affil{University College London, Mullard Space Science Laboratory, Holmbury St. Mary, Dorking, Surrey, RH5 6NT United Kingdom}
\and
\author{P.A. Caraveo}
\affil{INAF - Istituto di Astrofisica Spaziale e Fisica Cosmica,
Via Bassini 15, I-20133 Milano, Italy}
\and
\author{G.F. Bignami\altaffilmark{1}}
\affil{Agenzia Spaziale Italiana, Via Liegi 26, I-00198 Roma, Italy}
\altaffiltext{1}{Istituto Universitario di Studi Superiori di Pavia, Via Luino 4, 27100
  Pavia, Italy}

\begin{abstract}
PSR  B0540-69 is the  Crab twin  in the  Large Magellanic  Cloud. Age,
energetic  and   overall  behaviour  of  the  two   pulsars  are  very
similar. The same  is true for the general  appearance of their pulsar
wind nebulae (PWNe).  
Analysis of Hubble Space Telescope images spanning 10 years
unveiled significant variability in the PWN surrounding  PSR  B0540-69,
with a hot spot moving at $\sim0.04$c. Such behaviour, reminiscent of the
variability  observed  in  the   Crab  nebula  along  the  counter-jet
direction, may suggest an alternative scenario for the geometry of the
system. The same data were used to assess the pulsar proper motion.
The null displacement recorded over 10 y allowed us to set 
a $3\sigma$  upper limit  of 290  km s$^{-1}$ to the pulsar  velocity.
\end{abstract}





\section{Introduction}

PSR  B0540-69  in the  Large  Magellanic Cloud  (LMC)  is  one of  the
youngest        pulsars        known       to        date\footnote{see
http://www.atnf.csiro.au/research/pulsar/psrcat/}  (characteristic age
$\tau \sim  1600$ years)  and one  of the very  few first  observed at
wavelengths other than radio.  It was discovered in X-rays by the {\it
Einstein  Observatory} \citep{seward84}, soon  detected to  pulsate in
the optical \citep{middleditch85}, but detected in radio only 10 years
later  \citep{manchester93}.  PSR  B0540-69 is  a fast  ($\sim50$ ms),
classical pulsar with a rotational energy loss very similar to that of
the Crab  ($\sim1.5\times10^{38}$ erg s$^{-1}$).   Thus, the detection
of  a polarized  plerion-like  structure \citep{chanan90}  came as  no
surprise, making it the most Crab-like of the Crab-like remnants.

With  the   first  high  resolution   optical  images  of   the  field
\citep{caraveo92} it  was possible  to identify the pulsar 
optical counterpart (V  $\sim$ 22.6),
disentangling it from the  surrounding structured plerion.
This picture was confirmed by  a snapshot Hubble Space Telescope (HST)
observation \citep{caraveo98} which clearly resolved, within $\sim4''$
from  the pulsar,  the plerion,  elongated in  the northeast-southwest
direction.

Using  early  Chandra  data, \citet{gotthelf00}  and  \citet{kaaret01}
performed a morphological  study of the plerion in  the X-ray band and
found a noticeable
similarity   (accounting  for  different   distance)  with   the  Crab
pulsar-wind nebula (PWN).
Furthermore,  \citet{gotthelf00} unveiled the  presence of  a brighter
PWN region south-west of the  pulsar. Somehow in analogy with the Crab
case,  they suggested  such region  to belong  to a  torus  around the
source,  since it appears  perpendicular to  a much  fainter structure
protruding from the pulsar and tentatively identified as a jet.

\citet{caraveo00}  performed  a multiwavelength  analysis  of the  PSR
B0540-69  PWN  morphology by  superimposing  Chandra  and HST  images,
finding
a good correlation  between the optical and X-ray  structures with the
PWN emission enhanced in both cases  SW of the pulsar, i.e.  along the
putative torus proposed by \citet{gotthelf00}.

More recently,  detailed studies of  the system, based on  both narrow
and wide-band HST observations \citep{serafimovich04,morse06}, further
strenghtened the similarity with the Crab owing to the presence of
a cage of  filamentary ejecta possibly originating, at  least in part,
in  a pre-supernova  mass ejection  phase  \citep[][]{caraveo98}.  The
complex interaction  between the PWN  and such envelope  is dominating
the plerion multiwavelength morphology,
as  confirmed by  \citet{petre07}, based  on  the analysis  of a  deep
Chandra observations.

Interestingly,  \citet{serafimovich04},  using  two HST  images  taken
$\sim$  4  years apart,  reported  a  tentative  pulsar proper  motion
measurement  of  $4.9\pm2.3$  mas  y$^{-1}$ (for  a  pulsar  projected
velocity  of  $1190\pm560$ km  s$^{-1}$),  aligned  with the  putative
southern  jet of  the  PWN. This  would  make PSR  B0540-69 the  third
pulsar, after the Crab \citep{caraveo99,ng06} and Vela \citep{caraveo01,dodson03} pulsars, with
a proper motion aligned with its  PWN jet and, possibly, with the spin
axis, which  would have important consequences for  pulsar kick models
as well for the studies of the pulsar/PWN interactions.

In this Letter,  we report the results of our  analysis of the extended
PSR B0540-69 HST archived dataset
which allowed us   to study possible morphological changes 
in the PSR B0540-69 PWN
and to assess the pulsar proper motion  over a longer time
baseline.


\section{HST data analysis and results}
\label{dataanalysis}

Recent HST observations of PSR  B0540-69 were performed on 2005, November
15$^{th}$ with  the Wide Field  and Planetary Camera 2  (WFPC2), using
the wide  band  F555W  (480 s) and  the medium band  F547M (1040
s) filters. Such data add up to WFPC2 observations collected in
1995 and in 1999 with  several different filters \citep[see][for a log
of the  observations]{morse06}.  
In order  to study the PWN variability as well as the
pulsar proper motion an accurate superposition of
multi-epoch data is required.
We decided  as a
first  step to  use only  the observations  performed using  the F555W
filter, having a better signal to noise. Moreover,
the  significant WFPC2  geometric  distortion, crucial  for a  correct
proper   motion   measurement,   has   been   accurately   mapped  for the
F555W filter
\citep{anderson03}.
The selected dataset, which also provides a time baseline of $\gtrsim$
10  years,  includes the  recent  2005  observation,  as well  as  the
original 1995, October  19$^{th}$ (600 s) one \citep[][]{caraveo98}.

The  data  were retrieved  from  the  ST-ECF
archive and  reprocessed
on-the-fly using the most appropriate reference files.  Data reduction
and  analysis was performed  using the  IRAF/STSDAS, Midas  and FTOOLS
packages. Individual frames collected  during each visit were combined
to remove cosmic rays hits, and averaged.  
Residual  cosmic ray  traces  were removed  using specific  algorithms
within Midas.

In  order   to  
register the frames,
we  followed the procedure we already  applied in several
previous   astrometric   works with  HST   \citep[see
e.g.][]{caraveo96,deluca00,mignani00,caraveo01}.  A relative reference
frame was  defined for  each image, selecting  a sample of  good (well
resolved,  not saturated,  not  extended,  not too  close  to the  CCD
border) reference  sources. In the  crowded field of PSR  B0540-69, 85
good sources  were identified. Their position was  evaluated fitting a
2-D  gaussian function to  their intensity  profile, with  a resulting
uncertainty of  0.02-0.06 pixel per  coordinate.  The position  of the
pulsar  optical counterpart  was evaluated  in the  same way,  with an
uncertainty of order 0.03-0.04  pixel per coordinate.  The coordinates
of the  reference stars and of  the pulsar were then  corrected for the
WFPC2 geometric distortion using the mapping of \citet{anderson03}, as
well as for the ``34$^{th}$ row'' defect \citep{anderson99}.

Next,  the 1995  reference grid  was assumed  as a  reference  and was
aligned  along  Right  Ascension  and  Declination  according  to  the
telescope roll angle. Then,  we computed the best plate transformation
(accounting for independent  shift and scale factor for  each axis, as
well  as for  a rotation  angle) between  the two  grids  of reference
stars.  We applied an iterative clipping routine, to discard reference
stars  yielding  larger  residuals.   After  rejecting  15  stars,  we
obtained a  very good frame  superposition, with rms  uncertainties of
0.06 pixel per coordinate.


\subsection{Proper motion of the pulsar}

The  2005 pulsar position  was translated to the  1995 reference
frame,  to compute  the pulsar  displacement over  the 10.1  year time
span.  Such a displacement turned  out to be of 0.01$\pm$0.09 pixel in
R.A. and of 0.02$\pm0.08$ pixel in Dec, i.e. statistically null.
Using the well calibrated WFPC2  plate scale, we set a $3\sigma$ upper
limit to the source proper  motion of $\sim1.7$ mas~yr$^{-1}$.  At the
known pulsar distance  (51 kpc), such a limit  corresponds to an upper
limit to the total pulsar velocity (projected on the plane of the sky)
of $\sim410$ km s$^{-1}$.

As a further check, we  included in our astrometric analysis the WFPC2
observations  performed through  the  F547M filter  on 2005,  November
15$^{th}$ and on  1999, October 17$^{th}$ (800 s).   Indeed, the F547M
filter (pivot wavelength 5483\AA, $\Delta \lambda=483$\AA \, FWHM) has
a  narrower bandpass  than the  F555W one  (pivot  wavelength 5439\AA,
$\Delta  \lambda=1228$\AA  \,  FWHM),  but  the  pivot  wavelength  is
essentialy the same. Thus,  the wavelength dependence of the geometric
distortion  should  not induce  any  bias  when  using the  correction
optimized for the F555W filter.   The analysis was performed as above,
using the same reference stars. The superposition accuracy to the 1995
reference grid turned out to  be accurate within $\sim0.08$ pixels per
coordinate,  (i.e.    only  slightly   less  accurate  than   for  the
superposition of the F555W data),
with  no evidence  for  systematic  effects.  Also  in  this case,  no
significant displacement was measured for the pulsar. Combining
F555W and F547M data yields
a tighter  $3\sigma$ proper  motion upper limit  of 1.2  mas y$^{-1}$,
corresponding to a projected velocity of $\sim$ 290 km s$^{-1}$.
Such  a  limit,  computed   using  a  well-tested,  robust  algorithm,
supersedes  the  result by  \citet{serafimovich04},  based  on a  much
shorter time baseline. 

\subsection{Variability of the PWN}

The epoch-to-epoch  coordinate transformation  was then used  to rebin
and superimpose the images, in order to search for possible variations
of the  PWN morphology.  The  resulting images were not  corrected for
the geometric distortion.  However, we  note that at the PWN position,
imaged at the center of the PC chip, the maximum distorsion correction
is  $\sim0.05$ pixel  per coordinate  \citep{anderson03},  i.e.  small
enough for our goals.

As  a first  step, we  checked consistency  of photometry  between the
observations performed  through the  same filter at  different epochs.
To this aim, we compared the count rates from 85 reference stars using
simple  aperture  photometry.    The  2005  F555W  observations  yield
systematically lower count rates, with an average 2005--to--1995 ratio
of  0.86 (0.07  rms), while  the  2005--to--1999 ratio  for the  F547M
observations  is   0.90  (0.08  rms).    Such  values  were   used  to
re-normalize the 2005 images.

\begin{figure}
\plottwo{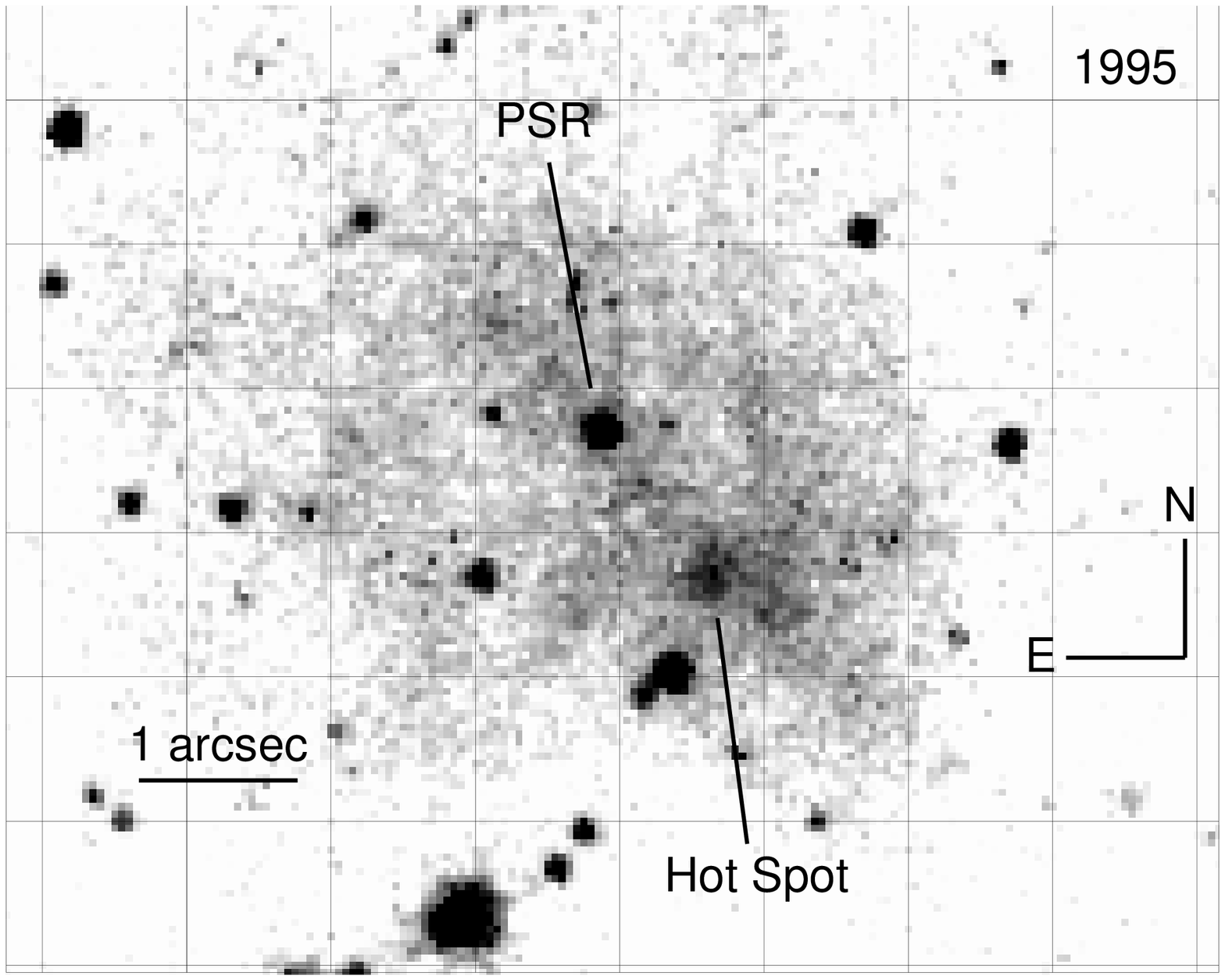}{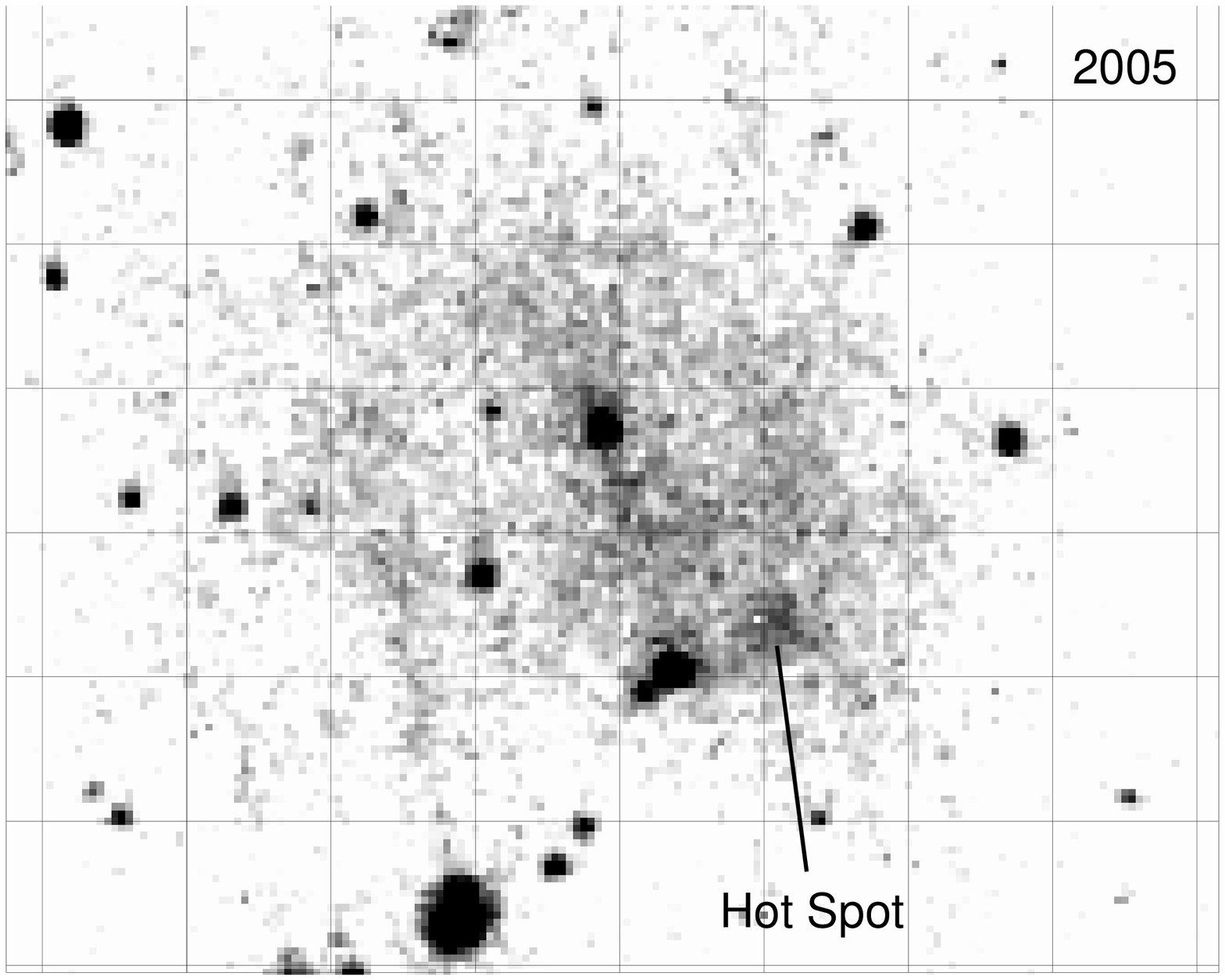}
\caption{The field of PSR B0540-69 as observed with
HST/WFPC2 through the F555W filter in 1995 
(600 s) and in 2005 (480 s). 
Pixel size is 0\farcs0455.
The pulsar optical counterpart, as well as the brightest feature
of the PWN (the ``Hot Spot'') are marked.
A grid (20 pixel spacing, corresponding to 0\farcs91)
is overplotted to better visualize the apparent
displacement of the ``Hot Spot''.   \label{1995vs2005raw}}
\end{figure}

We started using the 1995 and 2005 F555W images, characterized by a better 
signal-to-noise.
A striking change  in the brightest portion of  the PWN is immediately
apparent when comparing the two images.   As noticed by
\citet{caraveo00} 
a definite surface  brightness maximum is seen in  the PWN, South-West
to the pulsar optical counterpart.  Such a feature, which we will call
the  ``hot spot'',  is resolved  (7.2 pixel  FWHM, or  
0.081 pc
at the LMC distance) and lies $\sim1\farcs1$ away from the
pulsar 
(0.27 pc)
in the 1995  image.  Inspection of
the 2005 image shows that  the hot spot is displaced by $\sim0\farcs5$
in the  SW direction with  respect to its  location in 1995.   This is
apparent from Fig.~\ref{1995vs2005raw}, where the two images
are compared.
Fitting a gaussian function to the hot spot profile, we estimated that
the      feature     moved     by      $0\farcs46\pm0\farcs02$,     or
0.11 pc
 at  the  LMC distance,  corresponding to  a
velocity of $\sim0.037$c assuming simple linear motion.

The apparent displacement of the  hot spot corresponds to a very large
local  variation in the  PWN surface  brightness.  Using  a $8\times8$
pixel aperture  ($\sim0\farcs36\times0\farcs36$) centered on  the 1995
hot spot  position, the count rate  is seen to  decrease by $25\pm2$\%
between 1995 and 2005 (the uncertainty does not account for systematic
errors in  the image renormalization).  We have computed the  ratio of
the  2005 to  the 1995  surface brightness  using the  same $8\times8$
pixel aperture, selecting 200 positions to cover the whole PWN (within
$\sim2\farcs5$  from the pulsar).   Such an  exercise proved  that the
region of the hot spot is by  far the most active part of the PWN. The
observed  r.m.s.   variability  on   the  above  200  PWN  regions  is
$\sim8$\%,  which  reduces  to  $\sim4$\% when  considering  only  the
brightest 100  regions. 
Of
course,  the  possible  systematics  involved  in  the  2005  to  1995
renormalization do not affect such a conclusion.

The hot spot is clearly seen
in the image collected in 2005 with the F547M filter, 
at a position consistent with 
the one apparent in the F555W filter.
The  ratio of  the  2005 F547M  and  F555W images  does  not show  any
significant feature  at the hot  spot position, which suggests   the hot
spot emission  to be dominated  by continuum. Indeed, considering  a 4
pixel radius aperture centered at the hot spot, we estimated the ratio
of   the   observed    background-subtracted   count   rates   to   be
$0.4\pm0.1$. This is fully consistent  with an expected value of 0.45,
evaluated                 with                the                WFPC2
ETC\footnote{http://www.stsci.edu/hst/wfpc2/software/}, assuming
a        power-law       spectrum       of        spectral       index
$\alpha=$1.6\footnote{\citet{serafimovich04}          in         their
spatially-resolved  study of the  PWN spectrum  used a  region (``area
2'') encompassing the hot spot.   Emission from such a region, largely
contributed  by  the hot  spot,  is consistent  with  a  power law  of
spectral index $1.6\pm0.4$.}.

\begin{figure}
\epsscale{.70}
\plotone{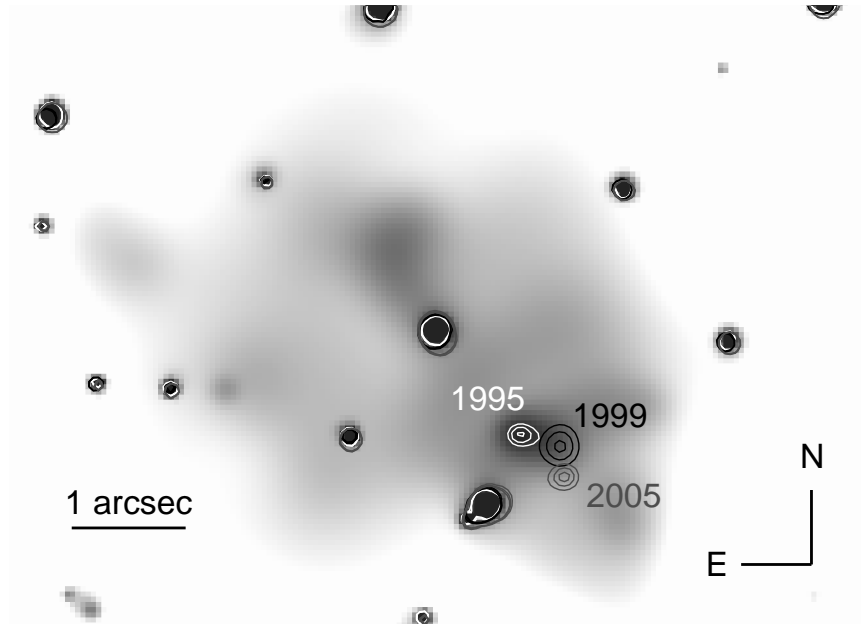}
\caption{The inner field of PSR B0540-69 as observed
in 1995 through the F555W filter is shown, after adaptive
gaussian smoothing of the HST/WFPC2 image. The {\em asmooth}
algorithm by \citet{ebeling06} has been
used, setting the minimum S/N threshold to 10 and using a maximum
smoothing kernel of 10 pixels. Isophotal contours to mark
the hot spot peak (corresponding to 99\%, 95\% and 99\%
of the maximum of the PWN surface brightness) are plotted 
in white. Contours generated using the same criteria for
the 1999 and 2005 images are overplotted. The displacement
of the hot spot is apparent. \label{multiepoch}}
\end{figure}

Thus,  we   can  use  the   1999  F547M  image  \citep[see   Fig.1  of
][]{serafimovich04} to  constrain the  position of the  hot spot  at a
third epoch.
Results are shown in Fig.~\ref{multiepoch}.
The  hot spot  peak  in 1999  lies  $\sim0\farcs38$ West  of its  1995
position, while in  2005 it is seen $\sim0\farcs33$  South of its 1999
position.  The hot  spot morphology is also seen  to vary, the feature
being more extended in 1999.
Such results argue against a simple outward
motion of the  feature and prove a dramatic variability  of the PWN in
the SW region.

The detection of large time variability in the PWN of PSR B0540-69
makes its similarity with the Crab Nebula even more compelling.
Thus, it seems natural to compare in some detail the optical phenomenology
of the two systems.
The  hot  spot  in  the  PSR B0540-69  PWN  is
definitely larger  and more distant  from the pulsar than  the bright,
highly variable  ``wisps'' seen  in the Crab  Nebula \citep{hester95,hester02}.
It is  somewhat reminiscent (as  for physical
dimensions, distance to the pulsar and temporal behaviour) of a large,
roughly arc-shaped  structure in the outer Crab  nebula, first noticed
in  the  optical  by   \citet{hester95}  because  of  its  outstanding
variability on a time scale of 6 years (see their Figure 12d).
Such  a  feature  is  also  prominent and  highly  variable  at  radio
wavelengths \citep[see Fig.2 of][]{bietenholz04}.  While the nature of
such feature is not understood,
it is  almost certainly related to  energy outflows  in the
counter-jet  channel  of the  Crab  PWN \citep{bietenholz04}.  Complex
interactions between the PWN  and the surrounding ejecta filaments are
also seen in such a region, roughly corresponding to the inner portion
of  the Norteastern  ``bay'' \citep{michel91,hester95}.   We retrieved
and inspected  HST/WFPC2 images of the  Crab collected in  1994 and in
2001 through the  F547M filter.  We found the outer  feature to show a
variability  consistent  to  the  one  reported  by  \citet{hester95},
corresponding to  a local surface brightness variation  of order 25\%,
very similar  to the  hot spot  of PSR B0540-69.   We also note that,
rescaled  at the  LMC distance,  the variability  in the  inner nebula
(wisps and torus) would be  difficult to detect, while the variability
of the outer structure would be outstanding.

Coming back  to PSR B0540-69, the hot  spot lies in the  region of the
PWN    tentatively   identified    as   an    equatorial    torus   by
\citet{gotthelf00}.  The large variability  of the feature, coupled to
the PWN asymmetry with respect to  the pulsar position, as well as the
comparison  with the case  of the  Crab, may  point to  an alternative
scenario  in which  the  northeast-southwest axis  corresponds to  the
direction of a  pulsar jet/counterjet.  This may also  be supported by
the  observation that  other  PWNe  do show  (in  X-rays) the  largest
variability along  the jet direction, with apparent  complex motion of
bright  blobs, although  on shorter  time scales  \citep[e.g. for PSR
B1509-58  and Vela,][]{delaney05,pavlov03}  as  well as  on a  smaller
physical  scale   \citep[in  Vela,][]{pavlov03}.   However,   no  firm
conclusions may be drawn based on current data.

\section{Conclusions}
With the discovery of significant variations in its PWN emission,
PSR B0540-69 shares one more characteristic with the Crab pulsar.
Moreover, our multi-epoch study of PSR B0540-69 yielded
a new assessment of the pulsar proper motion, setting an upper limit of 
290 km s$^{-1}$. 

Upcoming WFPC2 observations of PSR B0540-69 will allow us to monitor the PWN 
morphology, while lowering the measurable velocity to $\sim220$  km s$^{-1}$. 
The new HST data will  be very important 
to shed  light  on the geometry and dynamic of the system.
High-resolution HST polarimetric mode observations, to be collected in
the same program,  will offer unvaluable  clues in order to  understand the
overall structure of the PWN and its complex interaction with the cage
of  filamentary ejecta.  PSR  B0540-69 will  possibly become  a unique
extra-galactic laboratory to study  and understand the variability and
evolution of young PWN systems.

\acknowledgments
This work has been partially supported by the Italian Space Agency (ASI)
and INAF through contract ASI/INAF I/023/05/0. ADL acknowledges an ASI 
fellowship.

\clearpage

\clearpage

\end{document}